# Introduction to gravity model for beginners

Luigi Capoani

## Abstract

This paper provides a didactic, beginner-friendly review of the gravitational model and its conceptual translation from classical physics into international economics. After a brief introduction, it begins by establishing the structural and mathematical parallels between Newton's law of universal gravitation and the economic gravity equation, demonstrating how economic mass (GDP) attracts commercial flows while geographic distance acts as spatial resistance. The paper then examines the deterministic philosophical framework necessary to apply rigid natural laws to collective human behavior. Finally, it traces the chronological evolution of the literature, mapping the historical divergence between the physics-rooted approach of early demographers and Walter Isard, and some of the utility-based econometric adaptations. Ultimately, the paper demonstrates that despite globalization, spatial friction remains a vital, measurable force in shaping global commerce and geopolitical interactions.

## Introduction

Isaac Newton's discoveries revealed that massive physical objects like planets, stars, and moons are locked in a mutual attraction, pulling on one another through a force that can be calculated with mathematical precision. For centuries, his *law of universal gravitation* (1687) stood as the definitive proof that the physical universe operates under strict, predictable, and measurable rules. However, a similar order can be found in human society. In international economics, where the global marketplace often appears to be a chaotic whirlwind of free will and unpredictable consumer choices, the *gravity model* uncovers a striking geometric order within macro-level data. This model shows that international trade is governed by a social equivalent of physical gravity, relying on two core pillars: the economic size or wealth of trading partners, acting as their "mass", and the geographic distance separating them, acting as spatial friction or resistance.

Designed as a didactic overview, this paper explores the historical transition of physical paradigms into the social sciences, tracing this evolution from 19th-century demographers who viewed humans as literal molecules of society, through the spatial field theories of Walter Isard, to the utility-based econometric formulations that currently dominate international trade policy and geopolitical conflict analysis.

## 1. Mass, Distance, and Determinism: How Physics Models Society

To understand how the economic gravity model operates, we must examine its origin in classical physics with Newton's (1687) law of universal gravitation, which states that any particle of matter in the universe attracts every other particle with an invisible, pulling force. This physical interaction is governed by an exact mathematical equality. The magnitude of this attractive force ($F$) between two distinct physical masses ($M_i$ and $M_j$) is directly proportional to the product of those masses and inversely proportional to the square of the distance ($D_{ij}^2$) separating their centers. This natural relationship is expressed as:

$$F = G\frac{M_i M_j}{D_{ij}^2} \quad (1)$$

where $G$ represents the universal gravitational constant that, within the physical universe, remains perfectly invariant across cosmic space and time, anchoring a reductionist framework for calculating mass interactions.

When translated into international economics, this framework is adapted metaphorically to map onto the economic landscape:

$$F_{ij} = C\frac{GDP_i \cdot GDP_j}{D_{ij}} \quad (2)$$

In this economic equivalent, $F_{ij}$ indicates the total value of trade flowing between country $i$ and country $j$, while $C$ acts as the empirical constant of proportionality. A nation's Gross Domestic Product ($GDP$) directly takes the place of physical mass ($M$), as larger economies generate more domestic supply and possess greater importing purchasing power, expanding their capacity to "attract" commercial flows. Conversely, geographic distance ($D_{ij}$) represents spatial friction—or resistance—capturing shipping costs, transit times, and institutional trade barriers, rather than absolute, unyielding cosmic coordinates. Notably, while the physical formula (1) relies on the square of the distance, traditional economic models (2) simplify the denominator to a single power, re-calibrating the fundamental mechanics of physics to fit social space (Capoani, 2024). Additionally, unlike the unyielding cosmic constant $G$, the economic constant $C$ represents a dynamic baseline that fluctuates based on systemic variables such as asymmetric trade agreements, global production trends, technological developments, or historical transformations (Capoani, 2023).

Applying rigid mathematical formulas to human behavior requires a specific philosophical framework: determinism. This began with the 17th-century Newtonian mechanics, which introduced mechanistic determinism via the "*Principia Mathematica*" (1687) and provided a universal laws-based framework for understanding natural phenomena through predictable mechanical components (Capoani, 2025). This paradigm, known as the Laplacian-Newtonian paradigm or Laplacian determinism, asserts that macroscopic systems can be represented within a deterministic phase space, where the instantaneous configuration at any given moment constitutes the product of antecedent states under governing equations of motion. Indeed, as famously summarized by Pierre-Simon Laplace (1814), the present state of the universe is the effect of its past and the cause of its future. By the 19th century, physics shifted toward energetic physics and thermodynamics through the contributions of Carnot, Joule, and Clausius, moving from purely mechanistic descriptions to the study of complex energy transformations, entropy, and principles of conservation. Late 19th-century neoclassical economists systematically adapted these new physical concepts into economic theory, employing natural analogies such as market equilibrium and mathematical optimization (Capoani, 2025). Rather than viewing social behavior as entirely unpredictable, this paradigm uncovers an underlying structural determinism within aggregate human actions: while individual, microscopic human behavior appears chaotic and driven by subjective choices or free will, an invariant geometric order emerges when these actions are aggregated into massive macro-level economic networks. Consequently, just as physicists use statistical mechanics to calculate the macroscopic properties of gases without tracing every chaotic molecule, social sciences, sociology, and economic theory utilize the deterministic laws of gravity to map macro-level trends.

## 2. From People to Commodities: The Evolution of Economic Literature

While standard textbooks frequently credit Jan Tinbergen (1962) with pioneering the *gravity model*, a comprehensive bibliographic and historical review uncovers a much older origin stemming from 19th-century demography and intentional multidisciplinary borrowing from classical mechanics (Capoani, 2023). This trajectory began when Henry Carey (1858) systematically applied Newton's *law of universal gravitation* to human social systems, linking it to initial industrialization, railway traffic networks, and migration patterns (Capoani, 2023). Carey introduced the unique philosophical concept that an individual operates as a localized component of a grander structure—that he called a “molecule of society”—governed by an inescapable force of social attraction, causing people to gravitate toward large urban centers in the same manner point-masses attract one another in space (Capoani, 2025). This concept received its first formalized mathematical layout from Ernst Georg Ravenstein (1885), whose *laws of migration* demonstrated that expanding industrial hubs act as concentrated points of absorption that pull aggregate populations toward their centers in a predictable manner that grows weaker in strict proportion to the geographic distance traversed from the point of origin (Capoani, 2023, 2024, 2025).

By the early 20th century, this analytical framework shifted from demographic movement to commercial geography and retail market catchments. William J. Reilly (1929) adapted these identical principles of social gravitation to consumer behavior and urban markets, formulating the *law of retail gravitation*. He demonstrated that a city's capability to pull distant buyers is directly proportional to its population size and inversely proportional to the square of the distance to the urban center. By calculating the precise geographic “breaking point” where the gravitational pull of two competing cities reaches equilibrium, Reilly mapped a boundary of buyer indifference, proving that market boundaries obey spatial geometries (Capoani, 2023). The total formalization of “social physics” as an independent discipline was achieved in the 1940s by John Q. Stewart: he formalized exact equations for “demographic force" and “demographic energy”—preserving the strict physical adherence to the inverse square of distance—and analyzed populations as interacting masses (Capoani, 2023, 2024, 2025). To align his physical analogies with human systems, Stewart (1941) introduced a specific “molecular weight” to capture differing productivity levels among nationalities, setting the average American as his baseline mass unit. He also leveraged vector attributes to build continuous charts of demographic attraction that closely resembled physical field diagrams (Capoani, 2023).

The definitive transition from demographic migration to international trade occurred through the work of economist Walter Isard (1954) (Capoani, 2024). Inspired by Stewart's socio-physics, Isard sought to overcome the strict boundaries of classical trade models, which generally ignored spatial variables and treated international commerce as if it occurred in a frictionless vacuum (Capoani, 2025). He argued that trade theory and industrial localization are conceptually identical, meaning that global commerce is inextricably tied to physical space and transportation expenses. To execute this theoretical blend, Isard established a gravitational template that modeled trade flows as part of a wider framework of spatial concentration, wealth distribution, and attraction forces (Capoani, 2024, 2025). Replacing standard two-country economic models with a multilateral gravitational field approach, he formulated the concept of "income potential" to map how the aggregate economic mass of a multitude of surrounding nations simultaneously exerts a pulling force on any single country, anchored by the philosophical conviction that distance variables operate deterministically across the social world as they do in the natural world (Capoani, 2023).

A major theoretical split occurred in 1962 when Jan Tinbergen and his team introduced their own gravity equation (Capoani, 2023, 2024). In contrast to Isard's econophysical focus on global magnetic structures, Tinbergen was primarily interested in creating a pragmatic empirical tool for statistical regression (Capoani, 2025): his model removed Isard's intricate spatial field systems, presenting a highly simplified bilateral formula designed to estimate trade values strictly between pairs of nations. To maximize its empirical fit, Tinbergen's group added various non-physical, qualitative metrics to the distance variable, directly embedding institutional elements—such as colonial ties, postcolonial relationships, and shared language—into the spatial friction parameter (Capoani, 2023, 2024). This framework was subsequently canonized by Hans Linnemann (1966), whose doctoral thesis formalized the equation into a general equilibrium structure by adding specific parameters for domestic supply elasticities, potential demand, and artificial trade barriers (Capoani, 2023).

This historical division fragmented the evolution of the gravity model into two separate academic paradigms that continue to debate the role of physical analogies in economics (Capoani, 2024): econophysics and econometric. Econophysics—following the tradition of Carey, Stewart, and Isard—treats the model as a literal application of physical structures to human systems (Capoani 2023, 2024). It prioritizes the structural traits of spatial economics and analyzes macro-level commerce as dynamic fields governed by invariant, deterministic laws of attraction, systemic magnetism, and spatial concentration (Capoani, 2024, 2025). On the other hand, econometric—rooted in Tinbergen's lineage—treats the gravity equation as a flexible statistical tool (Capoani, 2023). James Anderson (1979) derived the gravity equation from consumer utility maximization and expenditure systems under perfect competition, anchoring the model to standard trade theory. This trajectory culminated with Anderson and van Wincoop (2003), who resolved traditional empirical anomalies—like McCallum's (1995) trade border puzzle—by introducing mathematical variables for "multilateral resistance". Their model proved that bilateral trade depends on trade barriers across the entire global economy rather than absolute physical distance, using the *iceberg trade costs* to show how spatial friction acts as an indirect tax on consumer utility (Capoani, 2023).

In contemporary literature, the model has been expanded far beyond its original boundaries to address institutional, historical, and geopolitical realities, showcasing its versatility across different scientific paradigms (Capoani, 2024, 2025). Modern econometric models use the gravity framework to quantify the economic damage of international conflicts, proving that political tension inflates the "economic distance" between nations, severing commercial ties more severely than a physical border would suggest (Capoani, 2024). Furthermore, Isard's spatial location frameworks have been integrated into peace economics, using gravitational fields to calculate how industrial co-location and dense cross-border commercial attraction can act as a structural deterrent against armed conflict (Capoani, 2023, 2024, 2025, 2025, 2026). While early social physics was explicitly deterministic and sought to uncover universal natural laws regulating human collectivities, modern mainstream economics completely inverted this logic. By tethering the gravity equation to individual rational choices and preference structures, modern econometrics decoupled the formula from its physical, deterministic genesis (Capoani, 2025). Today, the label "gravity" is frequently maintained in econometric studies for its suggestive and historical value (Capoani, 2023, 2024).

## Conclusions

Initiated as a direct transposition of the Newtonian gravitational paradigm to aggregate human dynamics, spatial economics originally operated under a strict deterministic framework. Early demographic models treated macro-level social networks as deterministic phase spaces, using classical mechanics to establish invariant equations for population absorption, retail boundaries, and demographic energy where spatial constraints functioned as absolute vectors. Under this approach, individual behavioral variations were treated as microscopic noise that neutralized upon aggregate observation, allowing macro-level social trends to be modeled with the same mathematical certainty applied to the statistical mechanics of physics. However, the evolution of the literature maps a major divergence in how spatial friction and mass are quantified and rationalized. The modern econometric paradigm stripped away these literal physical analogies, shifting the model's analytical foundation from natural laws to microeconomic choice theory. Furthermore, the transition from absolute physical distance to relative trade barriers—parameterized through multilateral resistance variables and *iceberg transport costs*—re-calibrated the model to account for dynamic, non-physical spatial frictions, transforming the denominator from a rigid cosmic coordinate into a function of endogenous policy and institutional friction.

Ultimately, the contemporary gravity framework maintains its high predictive capacity precisely because it bridges these two historical paths. While modern applications have decoupled the equation from its deterministic physical genesis in favor of flexible behavioral estimation, the resulting empirical regressions consistently validate the persistence of spatial boundaries. Whether formalized through the lens of econophysics as an invariant geometric field or parameterized through advanced econometrics to capture institutional disruptions and geopolitical margins, the structural interaction between economic mass and spatial resistance remains a mathematically robust, quantifiable reality of global market architecture.